\documentclass[twocolumn,showpacs,superscriptaddress,preprintnumbers,amsmath,amssymb]{revtex4}
\usepackage{graphicx} 
\usepackage{dcolumn} 
\usepackage{bm}
\usepackage{color}
\usepackage{times}
\begin{document}
\color{black}
\title{Dispersive Manipulation of Paired Superconducting Qubits}
\author{Xingxiang Zhou}
\affiliation{Department of Electrical and Computer Engineering,
University of Rochester, Rochester, NY 14627}
\affiliation{Department of Physics and Astronomy, University of
Rochester, Rochester, NY 14627}
\author{Michael Wulf}
\email{mikewulf@pas.rochester.edu}
\affiliation{Department of
Physics and Astronomy, University of
Rochester, Rochester, NY 14627} 
\author{Zhengwei Zhou}
\affiliation{Key Laboratory of Quantum Information, University of
Science and Technology of China, Chinese Academy of Sciences,
Hefei, Anhui, 230026, China}
\author{Guangcan Guo}
\affiliation{Key Laboratory of Quantum Information, University of
Science and Technology of China, Chinese Academy of Sciences,
Hefei, Anhui, 230026, China}
\author{Marc J. Feldman}
\affiliation{Department of Electrical and Computer Engineering,
University of Rochester, Rochester, NY 14627}

\date{March 9th, 2003}

\begin{abstract}
We combine the ideas of qubit encoding and dispersive dynamics to
enable robust and easy quantum information processing (QIP) on
paired superconducting charge boxes sharing a common bias lead. We
establish a decoherence free subspace on these and introduce
universal gates by dispersive interaction with a LC resonator and
inductive couplings between the encoded qubits. These gates
preserve the code space and only require the established local
symmetry and the control of the voltage bias.

\end{abstract}


\maketitle

Superconducting nano-circuits consisting of charge boxes (CB)
\cite{Charge} are among the most promising candidates for a
quantum computer. Coherent control of a single charge qubit
\cite{Nakamura}, long decoherence time \cite{Vion} and, more
recently, coupled two qubit systems \cite{Pashkin02} have been
demonstrated. But despite this encouraging experimental progress,
there are serious difficulties with superconducting QIP which may
appear insurmountable. The first is the severe decoherence
experienced by these macroscopic qubits, which are coupled to a
large number of degrees of freedom in their environment including
control circuitry \cite{Leggett87}. A few methods have been
employed to reduce the decoherence \cite{ChargeEcho,Caspar}, but
they usually require sophisticated manipulation or significant
overhead in the control circuitry. The second major difficulty
comes from the imperfect control realizable in solid state
systems. Specifically, one finds it difficult to achieve
controllable couplings between superconducting qubits, since the
commonly used hard-wired inductive or capacitive couplings are
untunable. Great effort has been exercised to realize controllable
couplings \cite{Makhlin99,You02,AV}. Schemes allowing to compute
with invariable couplings were also studied
\cite{Zhou02,Benjamin02}. Others have recently discussed to use an
LC circuit to actively mediate the interaction between
superconducting qubits \cite{Falci02,Blais02}.

The requirement to reduce decoherence and the desire for the
easiest manipulation apply to all QIP implementations.
Unfortunately, it is not always easy to accomplish both - actually
the goals are often contradictory - since reducing decoherence may
require extra complication in the manipulation.


In this work we show how to achieve both goals. Using a closely
placed pair of charge boxes (PCB) sharing a common bias lead as
the logic qubit, we can encode information in a fashion immune to
collective noise, which is the dominating decoherence source in
our setting. We introduce LC resonators inductively coupled to the
PCBs whose virtual excitations allow us to manipulate the PCB
dispersively; all interactions will be off resonance, without
energy transfer \cite{Haroche, Scully}, and thus a logical qubit
stays within its encoding space even during manipulation. By
inductively coupling the CBs and taking advantage of dispersive
dynamics again, controlled phases can be induced between logical
qubits.

Combining dispersive dynamics and encoding offers a new method for
QIP. It overcomes the major difficulties of superconducting CBs in
a realistic and very simple fashion.
The only control required after initialization is to change the
voltage bias of the PCB.
%
Another advantage of our method is that it relies only on noise
symmetry over short distance and so this is a realistic technique
for scalable QIP over large systems.
Though we focus on superconducting QIP in this paper,
the general principle of dispersive manipulation of encoded qubits
will be valuable for many other QIP systems.

{\em Charge qubits and the dominating noise. --} In its simplest
form, the charge qubit is just a superconducting island voltage
biased through a Josephson junction. The Hamiltonian for the CB
system is $E_c(n-n_g)^2-E_J\cos\varphi$ \cite{Charge}, where the
charging energy $E_c=(2e)^2/2C_t$, $C_t$ being the total
capacitance of the island, is much greater than the Josephson
energy $E_J=I_c\Phi_0/2\pi$, and $\varphi$ is the phase drop
across the junction. When biased close to $n_g=C_gV_g/2e=1/2$
($C_g$ is the gate capacitance), it provides an effective
two-state system which can be used as a qubit. In the spin 1/2
notation,
\begin{equation}
H_{CB}=\frac{1}{2}B^z\sigma^z-\frac{1}{2}B^x\sigma^x,
\label{eq:H_CB}
\end{equation}
where the spin up or down states correspond to n=0 or n=1 excess
Cooper-pairs on the CB. The effective field $B^z=E_c(2n_g-1)$ can
be tuned by changing the gate voltage $V_g$. In order to control
$B^x=E_J$, we use a flux biased small dc-SQUID to replace the
junction, whose critical current is maximal $(I_c=I_c^0)$ when the
flux bias is off and vanishes $(I_c=0)$ when it is
$\Phi_0/2=\frac{h}{4e}$.

\begin{figure}[h]
    \centering
    \includegraphics[width=3in, height=2in]{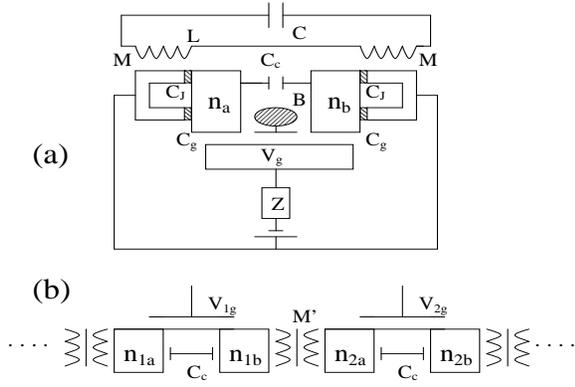}
    \caption{
    (a) A pair
    of closely spaced CBs sharing a common lead used as
    the encoded qubit. 
    The SQUIDS of both CBs are inductively coupled to an
    LC resonator whose virtual excitation allows the two charge
    boxes to exchange energy quanta. The circuit noise and background charge fluctuations are
    represented by $Z$ and $B$ respectively.
    (b) An inductively coupled PCB array. For
    clarity the PCBs are schematically represented by two boxes,
    the inductive coupling is understood to be between the
    dc-SQUIDS of the PCBs.
    }
    \label{fig:Charge}
\end{figure}

The dominating decoherence sources in this system are circuit
noise in the voltage bias \cite{Charge} and charge fluctuations in
the background (known as ``charge noise'')
\cite{Zorin96,Paladino02}, as indicated in Fig. \ref{fig:Charge}
(a). The circuit noise is described by the well known
``Spin-Boson'' model \cite{Leggett87}. The charge noise is less
well understood, but it is now generally believed to be caused by
fluctuations of the impurity charges in the substrate
\cite{Zorin96}. Taking into account the noise sources, we have the
total Hamiltonian for the system as $H=H_{CB}+H_Z+H_B+H_{int}$,
where $H_Z$ and $H_B$ are the Hamiltonian of the circuit
fluctuations, modelled as a collection of harmonic oscillators
\cite{Makhlin01}, and the
Hamiltonian of the background charge. 
Since these noise perturb the voltage bias of the CB, the
interaction Hamiltonian $H_{int}$ has the form
$H_{int}=\sigma^z\hat{E}^Z+\sigma^z\hat{E}^B$,
where $\hat{E}^Z$ and $\hat{E}^B$ are environment operators (with
the coupling strengths included).
Here we focus on the fluctuations in the voltage bias, the
dominating source of decoherence and neglect noise in the $B^x$
field (the critical current), as practised customarily
\cite{Leggett87, Makhlin01, Paladino02}. This is because charge
qubits are insensitive to flux noise and the effect of the
fluctuation in the $I_c$ suppression field is secondary to the
bias voltage variations discussed above. Detailed treatment of the
two voltage noise sources and their influence on the charge qubit
system can be found in the literature \cite{Paladino02,Makhlin01}.
For our purpose, the nature of the environment and specific form
of $\hat{E}^Z$ and $\hat{E}^B$ are not essential, therefore we do
not go into detail here.

{\em Paired charge boxes and DFS encoding. --} As shown in Fig.
\ref{fig:Charge} (a), we use two capacitively coupled identical
CBs ($a$ and $b$) with a common bias-lead as an encoded qubit. The
small capacitive coupling, $C_c\ll C_t$, is not essential for the
encoding, but necessary for the encoded two qubit gates. The
Hamiltonian of the PCB system is $H_{PCB}=\sum_{i=a,b}
\frac{1}{2}(B^z\sigma_i^z-
B_i^x\sigma_i^x)-\gamma\sigma_a^z\sigma_b^z$, where
$\gamma=\frac{C_c}{2C_t}E_c$ is the coupling energy.

Since the two CBs share the same lead, obviously they are biased
at the same voltage and they will experience the same circuit
noise. In addition the nano-scale charge islands are put
very close to each other. 
Therefore, they will experience the same charge fluctuations too
(more discussion on this point will be given later). Hence the CBs
experience ``collective decoherence,'' meaning the noise sources
couple symmetrically to them, which naturally gives rise to
``decoherence-free encoding.'' For a review of decoherence-free
subspace, see \cite{Kempe01} and references therein. Here we have
the simplest case of the DFS, with
$|0\rangle=|\downarrow_a\uparrow_b\rangle$ and
$|1\rangle=|\uparrow_a\downarrow_b\rangle$ as the decoherence-free
logical states. The way this works can easily be seen: as a
consequence of the collective decoherence the coupling between the
PCB and the environment is
$(\sigma_a^z+\sigma_b^z)*(\hat{E}^B+\hat{E}^Z)$,
 which annihilates the
two logical states given above. Therefore the PCB system will not
get entangled with the environment if it is initialized and kept
in the DFS. Physically, the encoded qubit's immunity to noise
stems from the fact that the CBs acquire random but opposite
phases.

To prepare the system in the DFS, we bias the PCB far off the
degeneracy point $n_g=1/2$. In the spin-1/2 picture, this
corresponds to applying a strong field in the $z$ direction. At
low temperatures, the spins will line up with the field, and the
PCB relaxes to the state $|\downarrow_a\downarrow_b\rangle$.
Keeping both $B^x$ off, we change $B^z$ (same for $a$ and $b$) to
$-2\gamma$. This cancels the bias of $a$ on $b$ making its total
$B^z_{tot}=B^z+2\gamma=0$. After that we turn on some $B_b^x$.
After a time $\pi/B_b^x$, the state of $b$ will become
$|\uparrow_b\rangle$, and the system is prepared in the DFS state
$|\downarrow_a\uparrow_b\rangle$. Now we turn off $B_b^x$ as well,
and from now on the $B^x$ fields for both CBs will remain off, by
biasing the dc-SQUIDS of the PCB at $\Phi_0/2$. Since $B^x$ fields
will remain off and need not be tuned after the initialization,
the leads tuning $I_c$ of the PCB could be heavily filtered to
keep out the noise once the system is initialized. Alternatively
we could make use of the noise free constant flux-bias techniques
such as that demonstrated in \cite{MooijAPL}. In practice it can
be difficult to suppress $I_c$ of the PCB precisely to 0 due to
the finite self inductance of the dc-SQUID. However as shown in
\cite{You01} if low self inductance dc-SQUIDS are used
($\frac{LI_c}{\Phi_0}\ll 1$) the $B^x$ field at $\Phi_0/2$ bias
point is negligibly small compared to $B^z$ fields used for
computation and can thus be safely dropped. Many schemes
\cite{Charge, Makhlin99, You02} rely on this fact too. One notices
that the logical states $|0\rangle$ and $|1\rangle$ are always
degenerate regardless of the voltage bias $n_g$, therefore there
is no evolution in the idle mode, regardless of the voltage bias
or noise in it. To readout the state of the PCB, a measurement of
its CB $a$ or $b$ will suffice, which is readily accomplished with
developed techniques \cite{Makhlin01}.

As seen above, in realizing DFS encoding with the PCB, we lose
considerable freedom in manipulating the system. First, in order
to guarantee symmetrical coupling to the circuit noise, the two
CBs share the same lead and hence they are always biased at the
same voltage. More importantly, we must ensure that operations on
the PCB do not drive the system out of the DFS, otherwise the
immunity to noise is lost \cite{Kempe01}. This is why we must keep
the $B^x$ fields for the CBs off: they flip the states of a single
CB and hence do not preserve the DFS. Therefore the only control
left is the voltage bias of the PCB, which clearly is insufficient
for universal QIP on the PCB. To deal with this difficulty, we
introduce a LC resonator inductively coupled to the PCB system:

{\em The LC resonator inductively coupled to the PCB. --} As shown
in Fig. \ref{fig:Charge} (a), we inductively couple the dc-SQUIDS
of the two CBs in the PCB system symmetrically to an initially
unexcited LC circuit. Even in the ground state its vacuum
fluctuations bias the dc-SQUIDS of the PCB off $\Phi_0/2$ making
charge tunnelling possible. The Hamiltonian for the PCB-LC system
is $H=H_{PCB}+H_{LC}+H_{coup}$, where $H_{LC}=\omega a^\dagger a$,
and $H_{coup}=-ig(a-a^{\dagger})\sigma_x$ with $
g=\frac{1}{2}MI_c^0\sqrt{\frac{\hbar\omega}{2L}}$.
%
Here $\omega$ and $L$ refer to frequency and inductance of the
LC-resonator; M is the mutual inductance between SQUID and
resonator. When the PCB and LC are far off resonance, the effect
of the LC can be neglected. On the other hand, when we tune the
bias of the PCB such that it is close to being in resonance with
the LC resonator, within the framework of Rotating Wave
approximation the above Hamiltonian becomes
%
 $H=H_{PCB}+H_{LC}-ig
\sum_{i=a,b}(\sigma_i^+a-\sigma_i^-a^\dagger)$,
where $\sigma^\pm=(\sigma^x\pm i\sigma^y)/2$ are the ladder
operators. Notice that $H_{PCB}$ contains a coupling term , in
contrast to the standard Jaynes Cummings Model which modifies the
dynamics significantly, as will be seen below.

Let $\delta=B^z-\omega$ be the detuning. If we let the PCB and the
(initially un-excited) LC resonator interact right in resonance,
i.e., $|\delta-2\gamma|\ll g\ll \omega$ or $|\delta+2\gamma|\ll
g\ll \omega$ \cite{Note}, state transfer occurs between the PCB
and the LC resonator \cite{Falci02, Blais02}. This is not allowed
in our scheme, since it will drive the PCB out of the DFS.
Besides, once the LC resonator is excited, we will have additional
decoherence due to the finite quality of the LC resonator
\cite{Falci02, Blais02}. Therefore, we only work in the dispersive
region, $g\ll |\delta\pm 2\gamma| \ll \omega$. In this case, the
PCB and the LC resonator cannot exchange energy because of the
large detuning. However the virtual excitation of the LC resonator
gives rise to an effective interaction between the two CBs
\cite{Zheng},
\begin{equation}
H_{eff}=\frac{g^2}{\delta+2\gamma} +
\frac{g^2}{\delta+2\gamma}(\sigma_a^+\sigma_b^-
+\sigma_a^-\sigma_b^+), \label{eq:H_eff}
\end{equation}
where the first term describes the Stark shift, which can be
neglected since it is the same for both logical states,
and the second term is the effective exchange interaction
caused by the exchange of a virtual photon. It preserves the DFS
(it changes $|\uparrow_a\downarrow_b\rangle$ to
$|\downarrow_a\uparrow_b\rangle$ and vice versa) and acts as a
logical $X$ gate on the PCB. Starting from
$|0\rangle=|\downarrow_a\uparrow_b\rangle$, letting the PCB evolve
under the effective Hamiltonian (\ref{eq:H_eff}) for a time
$t=\frac{\pi(\delta+2\gamma)}{g^2}$ or $\frac{t}{2}$, we can swap
the states of $a$ and $b$ or generate a maximally entangled state
between them.
%
The LC resonator is initially in the vacuum state and will not be
excited due to the dispersive interaction with the PCB; therefore
unlike in previous schemes \cite{Falci02, Blais02}we are not
subject to decoherence caused by its finite quality. Also, notice
that our dispersive scheme is fundamentally different from the
previous method of using the virtual excitation of a large LC
circuit capacitively coupled to all the qubits \cite{Makhlin99},
in which the LC-frequency is much larger than the CB-energies and
the coupling strength is tuned by changing the $E_J$'s of the
qubits. In our scheme the two energies are close (though the
detuning is large) and the $E_J$'s are always off; the coupling
strength is controlled by simply changing the detuning. Our scheme
takes advantage of the quantum exchange effect (Eq. (2)) assisted
by a virtually excited quantum state of the LC-qubit system
\cite{Zheng,Haroche01}. It offers noise protection with simplest
operation using very realistic parameters (see below), which was
not available in previous schemes.

To realize $SU(2)$ on the PCB, we still need a phase gate. This
can be accomplished by using another LC resonator, not depicted in
Fig. \ref{fig:Charge} (a) and at a frequency $\omega'$ very
different from $\omega$, inductively coupled to the dc-SQUID of
only $a$ (or $b$) in the PCB. Then when we tune the voltage bias
of the PCB such that it interacts with this LC resonator
dispersively ($g\ll |\delta'\pm 2\gamma|\ll \omega'$), a phase
gate will be obtained due to the Stark shift
$\frac{g^2}{\delta+2\gamma}|1\rangle \langle 1|$. The effect of
the previous LC resonator can be neglected because it is far off
resonance with the PCB. However for the purpose of universal QIP
this 
second LC resonator is not absolutely necessary
\cite{Shi02}, 
as will be shown below.

{\em Inductively coupled PCB arrays. --} To realize universal QIP
on the PCBs, we need a scheme to couple them. We notice that
different PCBs will experience different noise as they are biased
by different leads. Those far apart are susceptible to different
charge noise too. So we only have what we call ``local'' DFS; this
only relies on noise symmetry over a few, here two, physical
qubits, as is inevitably the only realistic case for scalable QIP,
and previously discussed methods \cite{Kempe01} do not apply.
Stringent restrictions are put on the two qubit coupling in order
to preserve the DFS. A capacitive coupling between $1b$ and $2a$,
for example, cannot be used, as this would cause the noise in
PCB1's lead to leak asymmetrically into PCB2. Furthermore neither
PCB may leave its DFS during its evolution.

Here we discuss a new approach that allows scalable QIP based on
local DFS.
We couple the PCBs
inductively, using a small mutual inductance $M'$ between their
dc-SQUIDS, as shown in Fig. \ref{fig:Charge} (b).
 As the dc-SQUIDS are biased
at $\Phi_0/2$ the coupling Hamiltonian is
$\lambda\sigma_{1b}^x\sigma_{2a}^x$ \cite{You01}, where the
coupling strength $\lambda=\frac{3}{4}M'I_{c,1b}^0I_{c,2a}^0$
(chosen much smaller than $E_J^0$, the unsuppressed Josephson
Energy of the SQUIDS).

Obviously the PCBs in the array in Fig. \ref{fig:Charge} (b) can
be initialized in their DFSs just as described before. Now, if all
the PCBs are biased at the same voltage, they will get out of
their DFSs due to resonant interactions by the above coupling
term. Therefore, we bias the odd numbered PCBs at the degenerate
point $n_g=1/2$, giving a $B^z=0$ and the even numbered PCBs at
some other point (but far off resonance with their LC resonators)
giving a large $B^z$. Because of the large detuning, $B^z\gg
\lambda$, the coupling Hamiltonian has no effect in the idle mode
\cite{Benjamin02}. One might worry that this different biasing
will introduce a difference between the phase frequencies of the
PCBs, but it is not the case since the DFS states always have the
energy $\gamma$ regardless of the voltage bias of the PCBs. When
we want to do a controlled phase gate between PCB 1 and 2, we tune
their biases near some common target value very different from
their previous values (such that they do not interact with other
neighboring PCBs) and the LC frequencies. Assuming the fields are
$B_1^z$ and $B_2^z$, we work in the dispersive region such that
the states of the PCBs do not change except that dispersive phases
are obtained due to the virtual transitions. For instance, the
energy of the state $|0_10_2\rangle$ will be shifted by
$\frac{\langle
0_10_2|\lambda\sigma_{1b}^x\sigma_{2a}^x|m\rangle^2}{E_{|0_10_2\rangle}
-E_{|m\rangle}}=\frac{\lambda^2}{4\gamma+\Delta}$, where
$|m\rangle=|\downarrow_{1a}\downarrow_{1b}\uparrow_{2a}\uparrow_{2b}\rangle$
is the virtually excited intermediate state and
$\Delta=B_1^z-B_2^z$ is the detuning. Other phases can be
calculated too, giving in the basis $|0_10_2\rangle$,
$|0_11_2\rangle$, $|1_10_2\rangle$, $|1_11_2\rangle$ an effective
Hamiltonian $diag\{\frac{\lambda^2}{4\gamma+\Delta},
\frac{\lambda^2}{4\gamma+(B_1^z+B_2^z)},
\frac{\lambda^2}{4\gamma-(B_1^z+B_2^z)},
\frac{\lambda^2}{4\gamma-\Delta}\}$, which reduces to $diag\{0, 0,
0, \frac{\lambda^2}{4\gamma-\Delta}\}$ in the dispersive region
$\lambda\ll |4\gamma-\Delta|\ll |4\gamma+\Delta|\ll |4\gamma\pm
(B_1^z+B_2^z)|$. This gives a CPHASE($\alpha$) gate $diag\{1, 1,
1, e^{-i\alpha}\}$, where
$\alpha=\frac{\lambda^2t}{4\gamma-\Delta}$ with $t$ being the
evolution time.

Notice that in the above implementation of the CPHASE gate, the
capacitive coupling $\gamma$ plays an essential role. As can be
easily checked the above procedure does not give an entangling
gate if $a$ and $b$ do not bias each other ($\gamma=0$). This is
because $|0_10_2\rangle$ and $|1_11_2\rangle$ would acquire
opposite energy shifts $\pm \lambda^2/\Delta$, due to the fact
that the energy differences between the initial and intermediate
states are opposite for these two cases. Thus the role of the
capacitive coupling $\gamma$ can be understood by an interesting
``parity argument:'' under the exchange of the states of $a$ and
$b$ for both PCBs, the energy mismatch (the denominator in the
perturbative calculation) due to the detuning $\Delta$ is odd.
This symmetry is broken by the coupling between $a$ and $b$, which
is even under this operation (as is obvious from the form
$-\gamma\sigma_a^z\sigma_b^z$).

Another point of interest is that, a phase gate on a PCB can be
implemented by using the CPHASE($\alpha$) and the X gate, as is
easily recognized by the identity
$e^{iZ_1\alpha}=e^{i\alpha}(X_2\cdot CPHASE(2\alpha))^2$.
Therefore, for the purpose of universal QIP on the PCB array the
LC resonator giving the phase gate is not absolutely necessary, as
long as the system has more than 1 qubit \cite{Shi02}. This is
potentially beneficial in reducing the hardware.

{\em Discussion. --} In the above we described our scheme of QIP
with PCBs based on encoded qubits and dispersive dynamics, which
requires only tuning the voltage bias of the PCBs. Our scheme can
prevent decoherence from collective noise. The circuit noise
obviously couples symmetrically to the CBs of the PCB. The charge
noise requires some caution, since its exact nature is still a
topic of debate \cite{Song95,Zorin96}. The early experiments in
\cite{Zorin96} clearly show that the charge noise on close by
islands are correlated. The conclusion drawn from this
observation, that the charge noise stems mostly from sources in
the substrate was further substantiated by \cite{Krupenin00}. This
has important consequences, because it suggests that it is
possible to engineer the environment for desired noise
configurations. Indeed, analysis in \cite{Zorin96} shows that high
noise correlations can be achieved for properly designed geometry
and layouts of the charge islands. Simple environment engineering
was already successful \cite{Krupenin00, Clark03} in various
contexts.

For our scheme to work, the CBs must be located within a distance
smaller than the wavelength of the background charge fluctuations,
so that they experience the same noise. This seems to be
realistic, since the advance in device fabrication allows to make
smaller structures, and more importantly the fact that the charge
noise originates from the substrate makes it possible to engineer
the environment for the desired noise symmetry
\cite{Zorin96,Krupenin00}. For instance, if we put the PCB on an
electrode instead of the substrate \cite{Krupenin00}, the charge
impurities will be located far away from the PCB, and thus couple
symmetrically to the CBs.
%
%

Though our scheme eliminates the effect of the collective charge
noise on the PCB, the decoherence time will be finite as there are
other non-collective noise in the system not dealt with by our
prescription, for instance dissipation due to the finite impedance
of the junction and the noise in its critical current. The effect
of the virtually excited states and the fluctuation of the
dispersive energies must be evaluated carefully too, though some
results were obtained previously \cite{Sorensen99,Gea02}.
Qualitatively, as shown by simple analysis based on Master
equations the number of operations allowed in our dispersive
scheme increases by $\Delta/g \gg 1$ (here, $\Delta$ the effective
detuning and g the coupling strength) as compared to the usual
scheme based on resonant Rabi manipulations, if the same coupling
strength $g$ is assumed \cite{Mike}. The PCBs do not experience
decoherence in the idle mode, which is favorable for a large
system in which only a fraction of the qubits undergo active
manipulation at the same time. Therefore, our scheme we can reduce
the error rate of the PCBs below the threshold for error
correction schemes \cite{QCbook} and thus make superconducting QIP
feasible. Detailed calculation of the decoherence time for a
realistic PCB system will be reported elsewhere \cite{Mike}.

Another practical concern is that the CBs in a PCB will not be
completely identical due to the imperfect fabrication. Because
only local symmetry is required, this problem is less significant
since fabrication variations tend to happen at large scales and
experiments show that closely spaced charge-boxes can be equal
beyond experimental resolution \cite{Zorin96}. In addition, the
error induced by non-identical qubits was shown to be higher order
in the symmetry breaking \cite{Lloyd}. Therefore, we conclude that
the technological problem of imperfect fabrication is already
solved to the extent needed for our scheme.

{\em Parameters. --} Finally, we give some parameters for the
experimental consideration. We use small CBs closely spaced with a
total capacitance $C_t\approx 0.16fF$ and charging energy
$E_c\approx 500GHz$. A mutual capacitance $C_c=5aF$ gives
$\gamma\approx 7.5GHz$. A mutual inductance $M=7pH$ between the
PCB and LC with $L=50pH$ and $\omega/2\pi=200GHz$ gives
$g=0.25GHz$ for $I_c=40nA$. Tuning the bias of the PCB close to
the LC frequency with a detuning $\delta\approx -12.5GHz$ results
in an exchange interaction with the strength
$\frac{g^2}{\delta+2\gamma}\approx 25MHz$, corresponding to a
period of $40ns$.
During the idle mode, we
bias the odd numbered PCBs at $B^z=0$ and the even numbered ones
at $B^z\approx 100GHz$. Since low self inductance dc-SQUIDS should
be used, we choose $M'I_c\approx 10^{-3}\Phi_0$ \cite{You01}, and
$M'=120pH$ gives the coupling strength $\lambda \approx 0.22GHz$.
Tuning the biases of the neighboring PCBs both to about $400GHz$
with a detuning $\Delta\approx 28GHz$ gives a CPHASE gate at the
rate $\frac{\lambda^2}{4\gamma-\Delta}\approx 25MHz$. The above
parameters are well within the reach of the current technology
\cite{Jena}.

In conclusion, we have discussed a technique for robust and easy
superconducting QIP. By combining the ideas of encoding and
dispersive manipulations, we protect the charge qubits from the
dominating decoherence and realize universal QIP on the encoded
qubits with minimal control. Besides the great potential of
solving the fundamental difficulties in superconducting QIP, we
expect the general idea of dispersive manipulation of encoded
qubits to be of interest to other physical systems, such as atomic
and other solid state systems \cite{Others}.


X. Zhou, M. Wulf and M. J. Feldman were supported in part by AFOSR
and funded under the DoD DURINT program and by the ARDA. Z-W. Zhou
and G-C. Guo acknowledge funds from NFRP, NNSF, and CAS of China.
%


\begin{references}
\bibitem{Note} These two conditions correspond to transitions to
$|\uparrow_a\uparrow_b\rangle$ and
$|\downarrow_a\downarrow_b\rangle$ respectively. Note here the
coupling energy $\gamma$ must be taken into account in determining
the in-resonance condition.
\end{references}
\end{document}